\documentclass{aa}

\newcommand{\kms}{km\,s$^{-1}$}
\newcommand{\masyr}{mas\,yr$^{-1}$}
\newcommand{\gaia}{\texttt{GAIA}}
\newcommand{\lamost}{\texttt{LAMOST}}

\usepackage{graphicx}
\usepackage{txfonts}
\authorrunning{ Zhong et al}
\titlerunning{ The Double Cluster populations}

\begin{document}

\title{The substructure and halo population of the Double Cluster $h$ and $\chi$ Persei}

\author{Jing Zhong \inst{1},
          Li Chen \inst{1,3},
          M.B.N. Kouwenhoven \inst{2},
          Lu Li \inst{1,3},
          Zhengyi Shao \inst{1,4}
          \and Jinliang Hou \inst{1,3} }

   \institute{Key Laboratory for Research in Galaxies and Cosmology,Shanghai Astronomical Observatory,Chinese Academy of  Sciences, 80 Nandan Road, Shanghai 200030, China,
   \email{jzhong@shao.ac.cn,chenli@shao.ac.cn}\\
   \and
    Department of Mathematical Sciences, Xi{'}an Jiaotong-Liverpool University, 111 Ren{'}ai Rd., Suzhou Dushu Lake Science and Education Innovation District, Suzhou Industrial Park, Suzhou 215123, China\\
    \and
    School of Astronomy and Space Science, University of Chinese Academy of Sciences, No. 19A, Yuquan Road, Beijing 100049, China \\
    \and
     Shanghai Key Lab for Astrophysics, 100 Guilin Road, Shanghai 200234, China}

   \date{}

% \abstract{}{}{}{}{}
% 5 {} token are mandatory

  \abstract
   {The \gaia{} DR2 provides an ideal dataset for studying the stellar populations of open cluster at larger spatial scale, as the cluster member stars can be well identified by their location in the multi-dimensional observational parameter space with the high precision parameter measurements.}
   {In order to study the stellar population and possible substructures in the outskirts of Double Cluster $h$ and $\chi$ Persei, we investigate using the \gaia{} DR2 data a sky area of about 7.5 degrees in radius around the Double Cluster cores.}
   {We identify member stars using various criteria, including their kinematics (viz, proper motion), individual parallaxes, as well as photometric properties. A total of 2186 member stars in the parameter space were identified as members.}
   {Based on the spatial distribution of the member stars, we find an extended halo structure of $h$ and $\chi$ Persei, about $6-8$ times larger than their core radii. We report the discovery of filamentary substructures extending to about 200~pc away from the Double Cluster.  The tangential velocities of these distant substructures suggest that they are more likely to be the remnants of primordial structures, instead of a tidally disrupted stream from the cluster cores. Moreover, the internal kinematic analysis  indicates that halo stars seems to be experiencing a dynamic stretching in the RA direction, while  the impact of the core components is relatively negligible. This work also suggests that the physical scale and internal motions of young massive star clusters may be more complex than previously thought.}
   {}


   \maketitle

%
%________________________________________________________________

\section{Introduction}

Almost all stars in our Galaxy form in clustered environments from molecular clouds \citep{2003ARA&A..41...57L,2010ARA&A..48..431P}. The study of stellar populations and morphology of open clusters, especially for young clusters, can help improve our understanding of many interesting open questions, such as the mode of star formation \citep{2009ApJS..181..321E,2015ARA&A..53..583H}, the dynamical evolution of cluster members \citep{2010ARA&A..48..431P,2014ApJ...787..107K}, and the dynamical fate of star clusters.

The $h$ and $\chi$ Persei Double Cluster (also known as NGC\,869 and NGC\,884, respectively) is visible with the naked eye, has been documented since antiquity. 
As one of the brightest and densest young open clusters containing large numbers of massive stars, the Double Cluster has been studied extensively  \citep[e.g.,][and references therein]{1937AnLei..17A...1O}. Over the last two decades, based on CCD photometry and spectrometry observation, general properties of the Double Cluster have been derived more accurately and reliably \citep{2002ApJ...576..880S, 2002A&A...389..871D, 2002PASP..114..233U, 2005AJ....130..134B,2010ApJS..186..191C,2013A&A...558A..53K}, with a resulting distance $d\approx 2.0-2.4$~kpc, an average reddening $E(B-V) \approx 0.5-0.6$, and an age of t$\approx 12.8-14$~Myr. The stellar mass function and mass segregation of $h$ and $\chi$ Persei were also interesting topics and explored in many works \citep[e.g.][]{2002ApJ...576..880S,2005AJ....130..134B,2010ApJS..186..191C, 2016MNRAS.457.1339P}, and provide an excellent opportunity to test models of cluster formation and early dynamical evolution. \citet{2002ApJ...576..880S} found evidence of mild mass segregation and a single epoch in star formation of the Double Cluster. In contrast, \citet{2005AJ....130..134B} claim finding strong evidence of mass segregation in $h$ Persei, but not in $\chi$ Persei. Moreover, the relationship between the Double Cluster and the Perseus~OB1 \citep{1992A&AS...94..211G} is still unclear, due to limitations of current observational data. Extensive observations suggest that a halo population contains a substantial number of member stars of  $h$ and $\chi$ Persei \citep{2007ApJ...659..599C, 2010ApJS..186..191C}. The halo region may extend well beyond 30~arcmin.  It is anticipated that a more extended and complete census of member stars within a radius of several degrees of the  $h$ and $\chi$ Persei cores will better reveal the star formation history of the larger region from within which the Double Cluster emerged \citep{2010ApJS..186..191C}.

The ambitious Galactic survey project \gaia{} survey aims to measure the astrometric, photometric and spectroscopic parameters of 1$\%$ of the stellar population in the Milky Way, and chart a three-dimensional Galactic map of the solar neighborhood \citep{2016A&A...595A...1G}. The recently released \gaia{} DR2 provides a catalog for over 1.3~billion sources \citep{2018A&A...616A...1G} with unprecedented high-precision proper motions (with typical uncertainties of 0.05, 0.2 and 1.2 mas\,yr$^{-1}$ for stars with $G$-band magnitudes $\le$ 14, 17 and 20~mag, respectively), parallaxes (typical uncertainties are 0.04, 0.1 and 0.7~mas, respectively) and also precise photometry (with typical uncertainties are 2, 10 and 10~mmag at $G=17$~mag for the $G$-band, the $G_{BP}$-band and the $G_{RP}$-band, respectively). The \gaia{} DR2 thus provides an ideal dataset for studying the stellar populations of $h$ and $\chi$ Persei at larger spatial scale, as the cluster member stars can be identified and distinguished from foreground and background objects based on their location in the multi-dimensional observational parameter space \citep{2018A&A...618A..93C,2018A&A...618A..59C}.

In this paper, based on the \gaia{} DR2 data, our main goals are to explore the large halo population and to chart for the first time the extensive morphology of $h$ and $\chi$ Persei. In Section~\ref{sec:data} we describe our method of membership identification. After performing the membership selection criteria, a total of 2186 stars are selected as candidate members.  In Section~\ref{sect:discussion}, we present the results for the halo population, and we present the discovery of the extended substructures in $h$ and $\chi$ Persei. A brief discussion about the Double Cluster formation and interaction is also present in Section~\ref{sect:discussion}.

\section{Data analysis} \label{sec:data}

As suggested by \citet{2010ApJS..186..191C}, to better reveal the large star formation history, the census region of young stars needs to extend to 5~degrees or more from the cores of $h$ and $\chi$ Persei. In our work, to explore the halo population at a larger scale, an extended region centered on $\alpha$=2$^{h}$17$^{m}$,$\delta$=57$^{\circ}$46$^{'}$ with a radius of 7.5~degrees was selected from the \gaia{} DR2 database. As the distance to cluster $h$ and $\chi$ Persei is about 2344~pc \citep{2010ApJS..186..191C}, this angular radius corresponds to a projected radius of about 300~pc, which is sufficiently large to cover the entire region of original giant molecular clouds from which the Double Cluster may have formed \citep{2015ARA&A..53..583H, 2016ApJ...821..125F}.

The initial catalog from \gaia{} DR2 has 9,195,956 sources, with $G$-magnitude ranging from 2.53 to 21.86~mag.
In the region encompassing the Double Cluster, this initial catalog contains the cluster members of $h$ and $\chi$ Persei, but the majority of sources are field stars and other background objects. Unlike field stars, star cluster member stars have similar bulk motions and distances, as all of the cluster stars were born in a Giant Molecular Cloud within a short period of time \citep{1993prpl.conf..245L}. This results in a relatively compact main-sequence in the color-magnitude diagram (CMD). The clumping of member stars in the multi-dimensional parameter space (proper motion, parallax and location in the CMD) can be used to perform the membership identification and largely exclude field stars and other background objects from the initial catalog.

\subsection{ Membership criteria}
\label{sec:criteria}
To exclude field star contamination, the expected proper motions and parallaxes as well as overall main-sequence distribution of $h$ and $\chi$ Persei have to be determined. The $h$ and $\chi$ Persei cluster shows a significant spatial central concentration in both cores. As a first step, stars within 10~arcmin from the centers of both cluster cores were selected as cluster member candidates, whose overall membership probabilities are substantially higher than that of stars located in the outer region \citep{2010ApJS..186..191C}. After this initial selection, total number of the initial member candidates in the core region is 14,031.

\begin{figure*}
   \centering
   \includegraphics[angle=0,scale=.3]{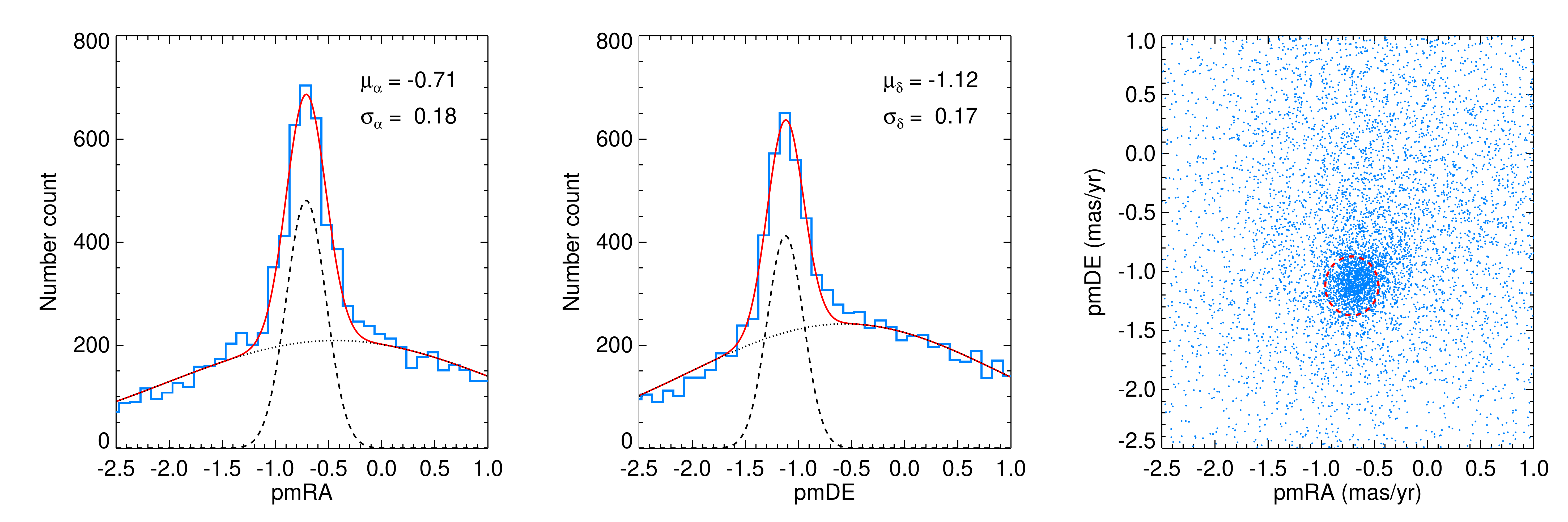}
  % \begin{minipage}[]{85mm}
   \caption{ Proper motion distribution of member candidates in the core regions ( $< 10'$ from the centres of $h$ and $\chi$ Persei).  The proper motion distribution in RA and DEC are shown in the left panel and central panel. Double-Gaussian profiles fit the distribution of members (dashed curve) and field stars (dotted curve) separately, while the solid red line shows the combined profiles. In the right panel, the red dashed line represent the membership criteria in the proper motion space, which includes the most probable member stars.}
%\end{minipage}
   \label{pm}
\end{figure*}

Subsequently, based on the member candidates in the core region, the average proper motion distribution ($\mu_{\alpha}$, $\mu_{\delta}$, $\sigma_{\mu}$) of members can be acquired by a double-Gaussian fitting in RA and DEC, respectively. The proper motion distributions of Double Cluster members is shown in Figure~\ref{pm}. The Double-Gaussian profiles were used to fit the proper motion distribution in RA (left panel) and DEC (central panel), shown as red solid lines here, while the black dashed and dotted curves represent member and field star distributions separately. The right-hand panel of Figure~\ref{pm} shows the proper motion distribution of initial member candidates in a focusing area. In proper motion space, we then use a circular region centered on the expected average proper motion ($\mu_{\alpha}$, $\mu_{\delta}$) = ($-0.71, -1.12$) mas/yr with radius 0.25~mas/yr  ($\sqrt{\sigma_{\alpha}^2+\sigma_{\delta}^2}$ ) as the membership criterion of the Double Cluster. After excluding possible field stars, which are located outside the criteria circle, 1294 core member candidates remain. 

\begin{figure*}
   \centering
   \includegraphics[angle=0,scale=.3]{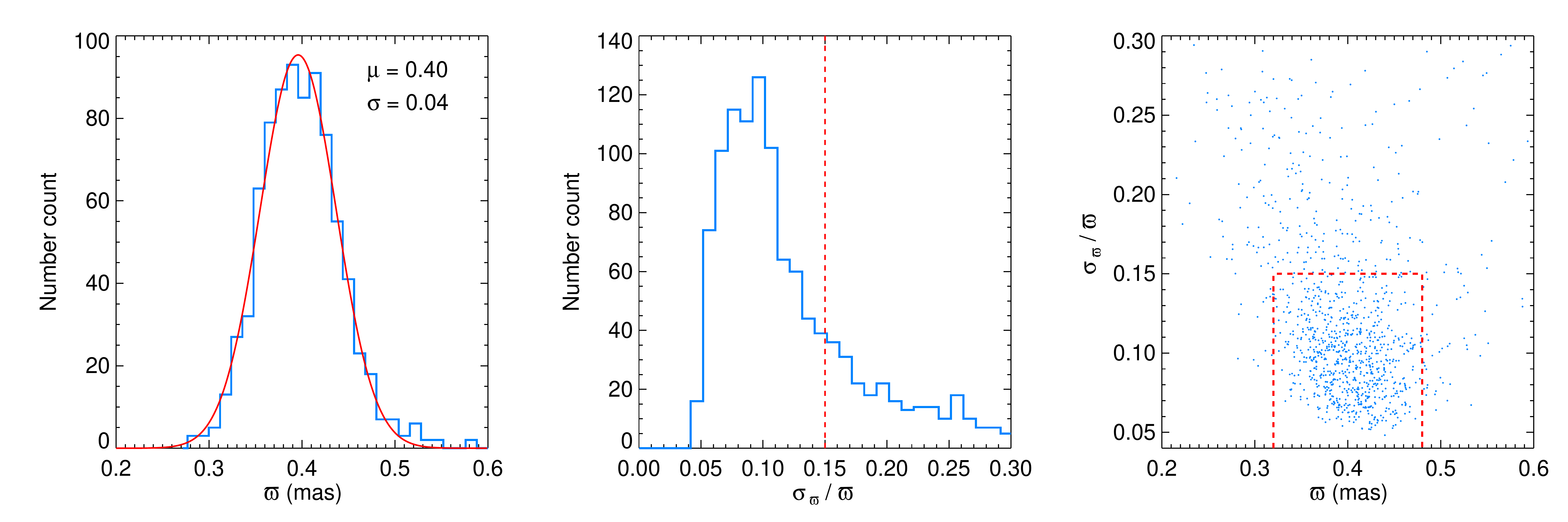}
  % \begin{minipage}[]{85mm}
   \caption{Parallax distribution of core member candidates which were selected by the proper motion criteria. In the left panel, the histogram of parallax $\varpi$ can be fitted by a single Gaussian profile. The distribution of relative deviation of $\varpi$ is shown in the central panel, with red dashed curve representing the acceptable maximum $\sigma_{\varpi}/\varpi$ boundary which including 68$\%$ candidates in the histogram. In the right panel, the parallax criteria is indicated with the red dashed curve.} 
%\end{minipage}
   \label{plx}
\end{figure*}

To further purify the member candidates, we used parallax criteria in Figure~\ref{plx} to exclude field stars from the 1294 remaining member candidates. In the left panel, the parallax distribution of those stars can be well fitted by a single Gaussian profile, which suggest the contamination by field stars is not significant, we  exclude stars with parallax $\varpi$ greater than 2$\sigma_\varpi$ of the Gaussian distribution. The histogram of relative deviation $\sigma_{\varpi}/\varpi$ is shown in the mid-panel with the peak located at about 10$\%$. To include the majority of member candidates, we select 15$\%$ as the criteria of relative deviation $\sigma_{\varpi}/\varpi$, which makes it containing 68$\%$ of remaining member candidates.  The $\sigma_{\varpi}/ \varpi $  vs. $\varpi$ diagram of the member candidates is shown in the right panel with blue dots. The parallax criteria is set as a region with $\sigma_{\varpi}/\varpi$ less than 15\% and $\varpi$ ranging from 0.32 mas to 0.48 mas (red dashed curve). After this parallax selection, the core member candidates retain 768 sources. 

\begin{figure}
   \centering
   \includegraphics[angle=0,scale=.25]{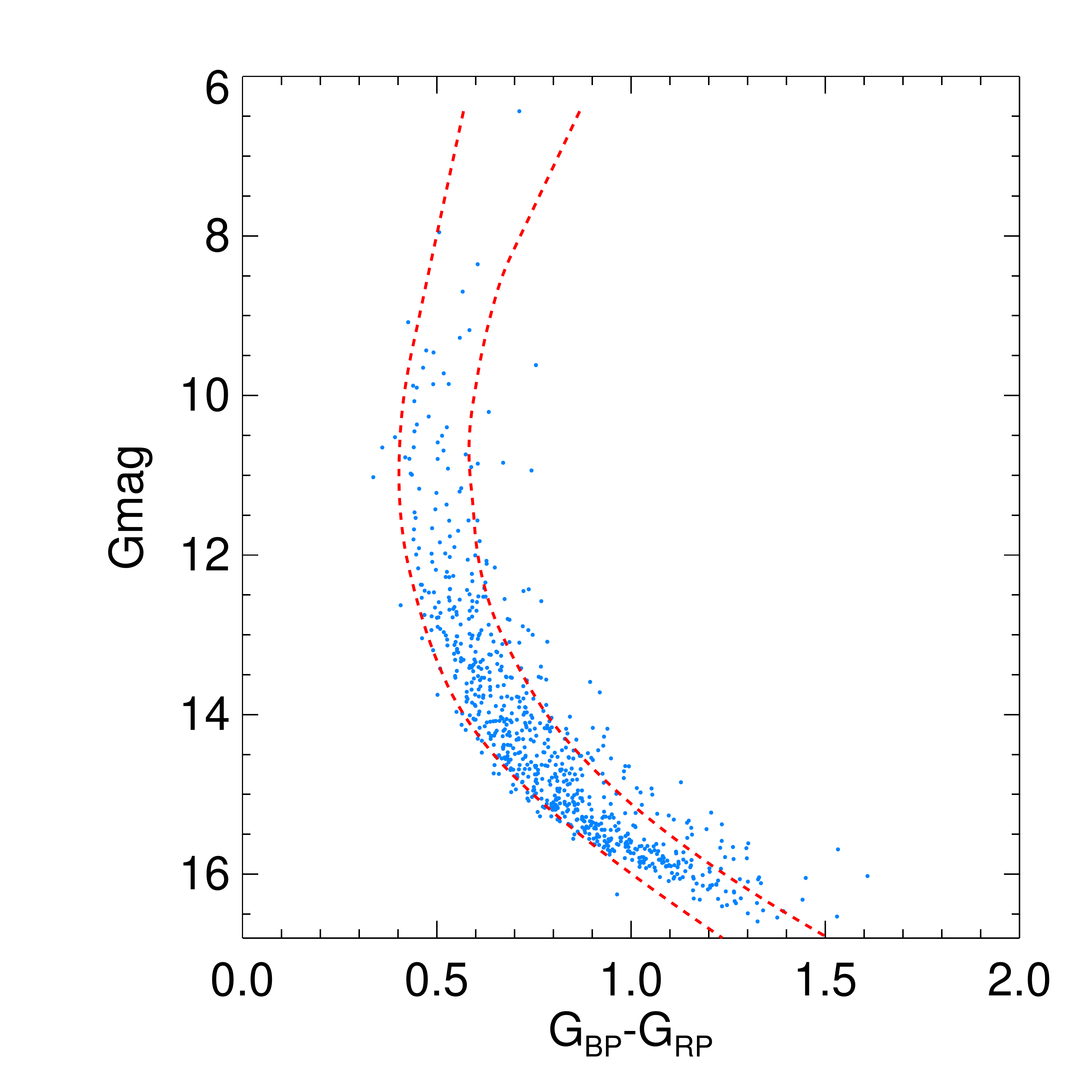}
  % \begin{minipage}[]{85mm}
   \caption{ $G$mag VS. $G_{BP}-G_{RP}$ color-magnitude diagram (CMD) for  astrometrically-selected possible member candidates. Two red dashed lines trace the 95$\%$ candidates distribution, and is regarded as the CMD criteria to further reducing field stars contamination. }   
%\end{minipage}
   \label{cmd}
\end{figure}

In the $G$mag VS. $G_{BP}-G_{RP}$ color-magnitude diagram, the limiting boundaries of  astrometrically-selected possible member candidates were also determined, as shown in Figure~\ref{cmd}. We divide  member candidates into several bins along with the $G$-band magnitude. For each magnitude bin, we use a Gaussian function to fit the stellar color distribution. The range between the lower and upper bounds contains 95$\%$ member candidates (2$\sigma$ clipping in color), and further reduce field stars contamination. To make the boundaries tracing the majority of isochrone distribution, we interpolate and smooth them from 6~mag to~16.5 mag in the $G$-band as is shown in Figure~\ref{cmd}. Finally, in the core regions, 637 core candidates are left after the CMD criteria procedure.

In summary, any remaining member candidate meets the following criteria: (i) the proper motion must be within a circular region which centered on ($\mu_{\alpha}, \mu_{\delta}) =(-0.71,-1.12$) mas/yr with radius 0.25 mas/yr (1$\sigma$); (ii) the parallax ($\varpi$) must be within the region (0.32, 0.48) mas, and its relative deviation of $\varpi$ less than 15$\%$; and (iii) the color-magnitude distribution ($G_{BP}-G_{RP}$, $G$mag) must be traced the majority of isochrone distribution of core members, as it shows in Figure~\ref{cmd}.

\section{ Results and Discussion}
\label{sect:discussion}

After establishing the membership criteria from the core member candidates of $h$ and $\chi$ Persei, we perform a member identification process to the initial catalog, which contains more than 9 million sources from \gaia{} DR2. After applying the selection criteria, 2186 stars are identified as member stars. It is noted that, our propose of identifying member stars is to explore the extended halo population and substructures of the Double Cluster at large spatial scale. Therefore, it is crucial to have a purer member sample in the trade-off between purity and completeness.

\subsection{ Spatial distribution }

\begin{figure*}
   \centering
    \includegraphics[angle=0,scale=0.4]{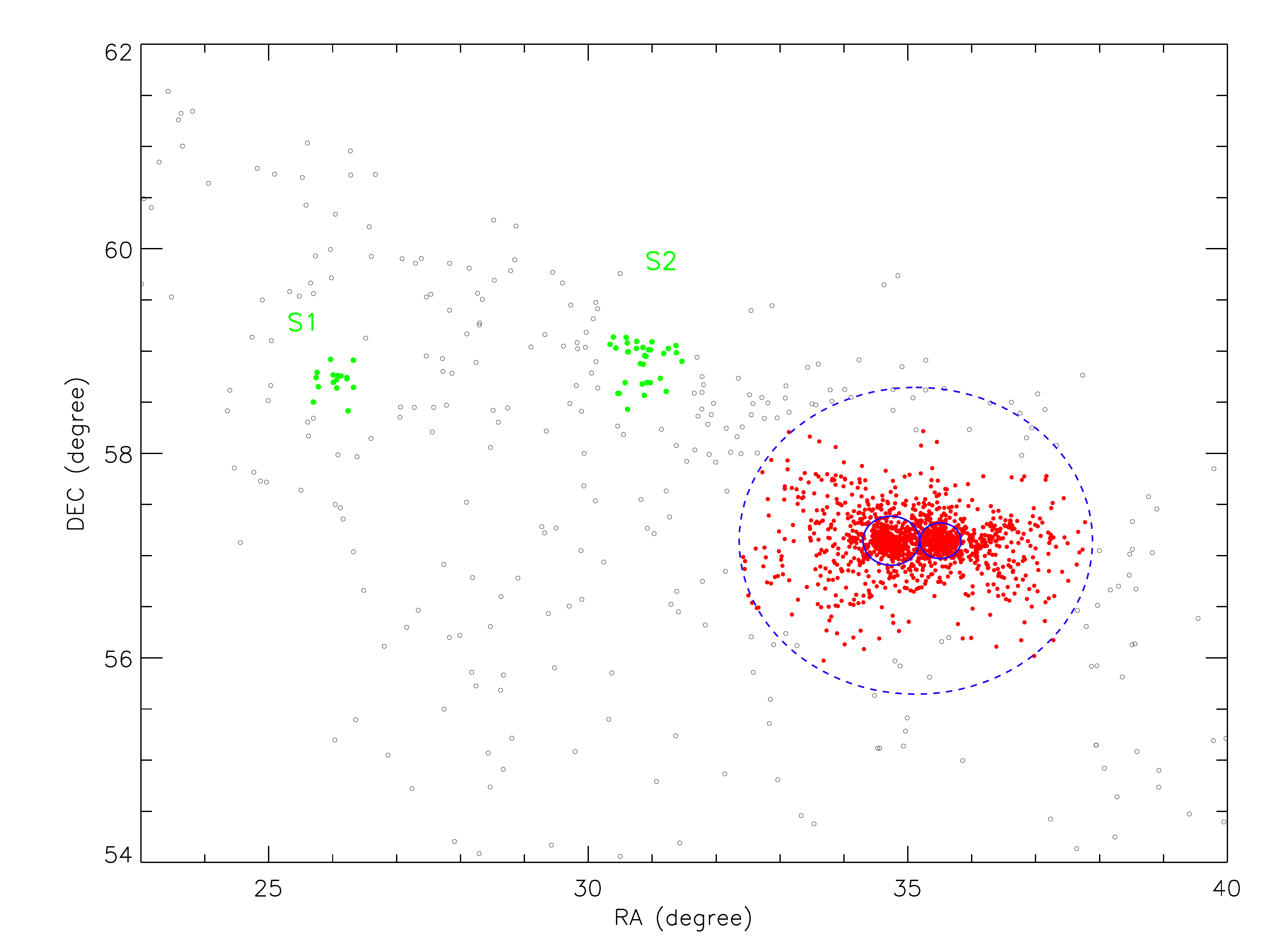}
  % \begin{minipage}[]{85mm}
   \caption{The spatial distribution of member candidates selected using the membership criteria. Based on the classification results from the DBSCAN clustering algorithm,   substructure components as well as halo and core components are marked as green and red dots respectively. The solid line represents the core region  from \citet{2013A&A...558A..53K}.  For comparison,  the extended halo region in this work is plotted with the dashed curve, with 1.5$^{\circ}$ in radius centered the Double Cluster centre, which is about $6-8$ times larger than the core radius of the Double Cluster.}
%\end{minipage}
   \label{dbscan}
\end{figure*}

Figure~\ref{dbscan} shows the spatial distribution of our member candidates. It is clear that in addition to the central core region of the Double Cluster, a significant number of member stars a spread out over a vast area, extending far beyond the $h$ and $\chi$ Persei boundary, as previously suggested by \cite{2013A&A...558A..53K,2002A&A...389..871D}.

To further probe the morphology of this extended halo and possible substructures, we adopt the clustering algorithm DBSCAN. This algorithm is widely used to define a set of nearby points in the parameter space as a cluster. The two DBSCAN parameters of define a set of cluster in our work are $\epsilon= 0.4$ and $minPts= 14$, which represent the radius and the minimum number of sources falling into the radius of the hypersphere, respectively  \citep[see more detail description in][and reference therein]{2018A&A...618A..59C}.

Figure~\ref{dbscan} also displays the DBSCAN classification results. Stars were classified into three groups. Red dots represent the group of cluster members located on the cluster core region and the adjacent halo region. The rest of two cluster groups are considered as cluster substructures and marked as green dots.

\begin{figure*}
   \centering
   \includegraphics[angle=0,scale=0.23]{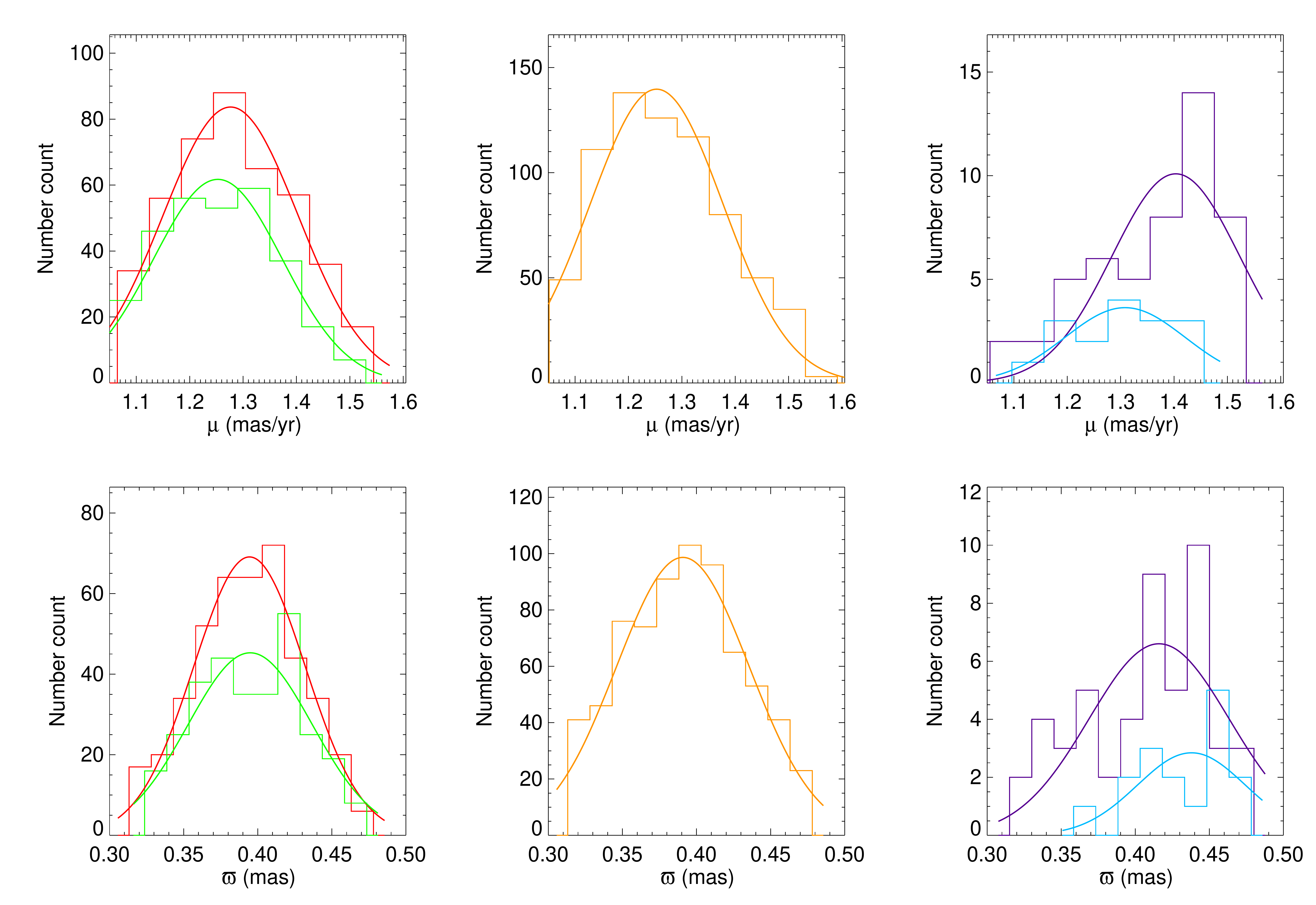}
  % \begin{minipage}[]{85mm}
   \caption{ Proper motion and parallax distribution of each component identified by the DBSCAN. Colored lines mark Double Cluster components, including  $h$ Persei (red), $\chi$ Persei (green), the halo population (yellow), S1 (blue) and S2 (violet). A Gaussian function is also applied to fit the histogram of each component. The fitted results are shown in Table~\ref{tab1}.}
%\end{minipage}
   \label{parm}
\end{figure*}

\begin{figure*}
   \centering
   \includegraphics[angle=0,scale=0.3]{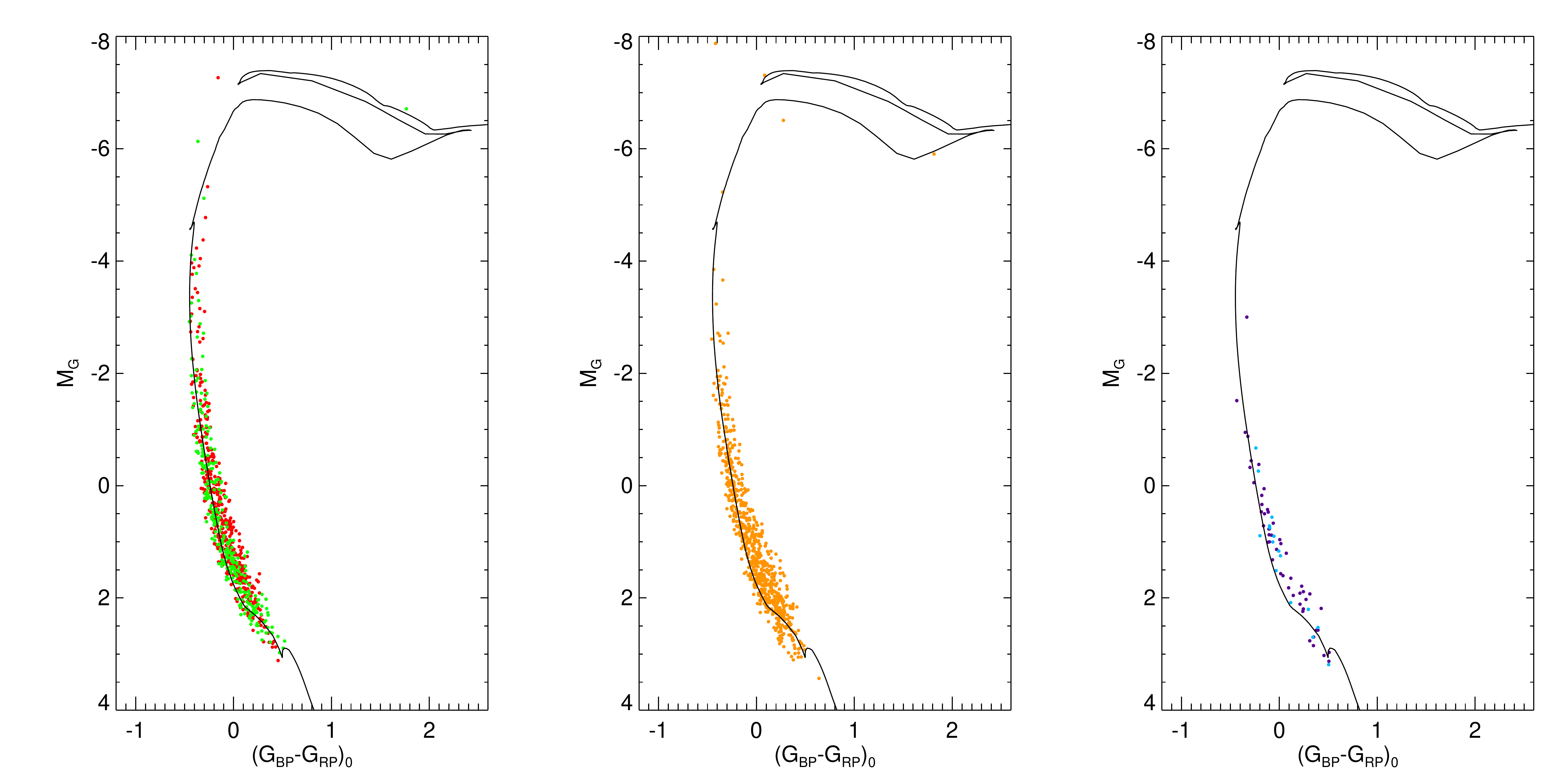}
  % \begin{minipage}[]{85mm}
   \caption{ De-reddened M$_G$ vs. $(G_{BP}-G_{RP})_0$ color-magnitude diagram for three major components which labeled by the DBSCAN. Colors are the same as in Figure~\ref{parm}. The Padova isochrone of 14~Myr is overlaid on the CMDs of three major components.}
%\end{minipage}
   \label{parm_hr}
\end{figure*}

\subsubsection{The halo population}
 
In much of the literature, the angular size of a star cluster is represented by the core radius, which is easily measured from the marked stellar overdensity features. \citet{2005A&A...438.1163K} suggest a corona radius as the actual radius of a cluster, which is defined as the radius where the stellar surface density equals  the density of surrounding field. In general, the corona radius is about 2.5~times larger than the core radius \citep{2005A&A...438.1163K}.
%In table~\ref{tab2}, we list the cluster radius estimation from literature results.

Based on our member sample selection, the Double Cluster possesses an extended low-density halo which has, to our knowledge, never been reported before. Its detection was possible only due to the unique capabilities of \gaia{}. In Figure~\ref{dbscan}, we plot the angular radius of the core region \citep[r1, from the catalog of][]{2013A&A...558A..53K}, as the blue circle. For contrast, we also plot a dashed blue circle, with 1.5$^{\circ}$ in radius around the cluster common center.  It is clear that our results shows a more extended region  for both $h$ and $\chi$ Persei, about $6-8$ times larger than their core radius. However, because of the incompleteness of the member sample, the halo radius of the Double Cluster is still regarded as a lower limit of the actual radius, even though the halo radius extends greatly beyond what was previously expected.

\begin{figure*}
   \centering
   \includegraphics[angle=0,scale=0.35]{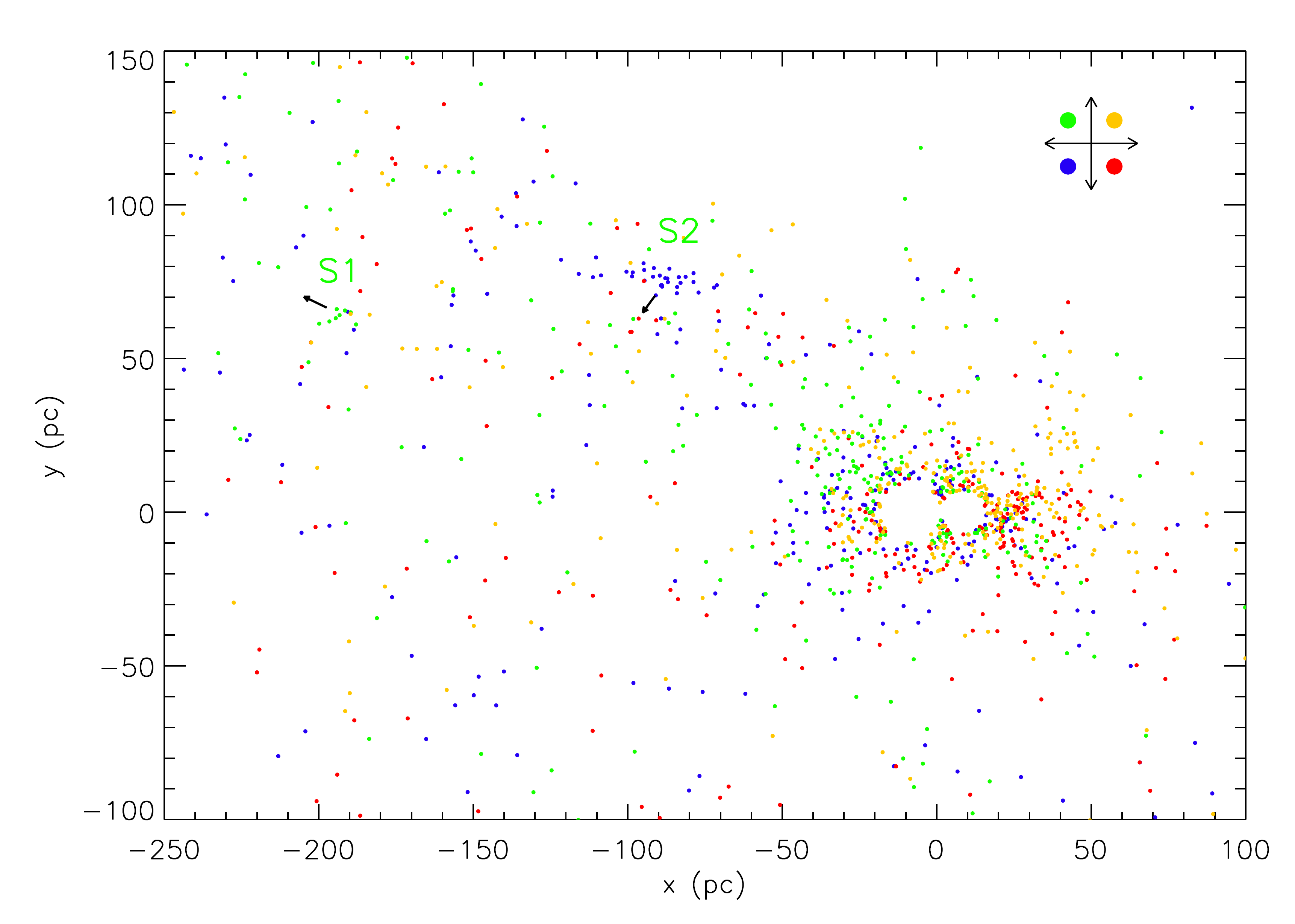}
  % \begin{minipage}[]{85mm}
   \caption{ Locations of member stars with different internal motion directions on the projected coordinate system, whose origin is located at the common center of the Double Cluster core. With the mean tangential velocity of core members as its zero point, stars are categorized into four samples according to their residual tangential velocity direction, as plotted with different colors. Arrows close to the label of S1 and S2 point out the average internal motion direction of substructure members. The almost orthogonal motion direction suggest the possibility of different dynamical formation histories of these two substructures. }
%\end{minipage}
   \label{qmap}
\end{figure*}

\begin{figure*}
   \centering
   \includegraphics[angle=0,scale=0.2]{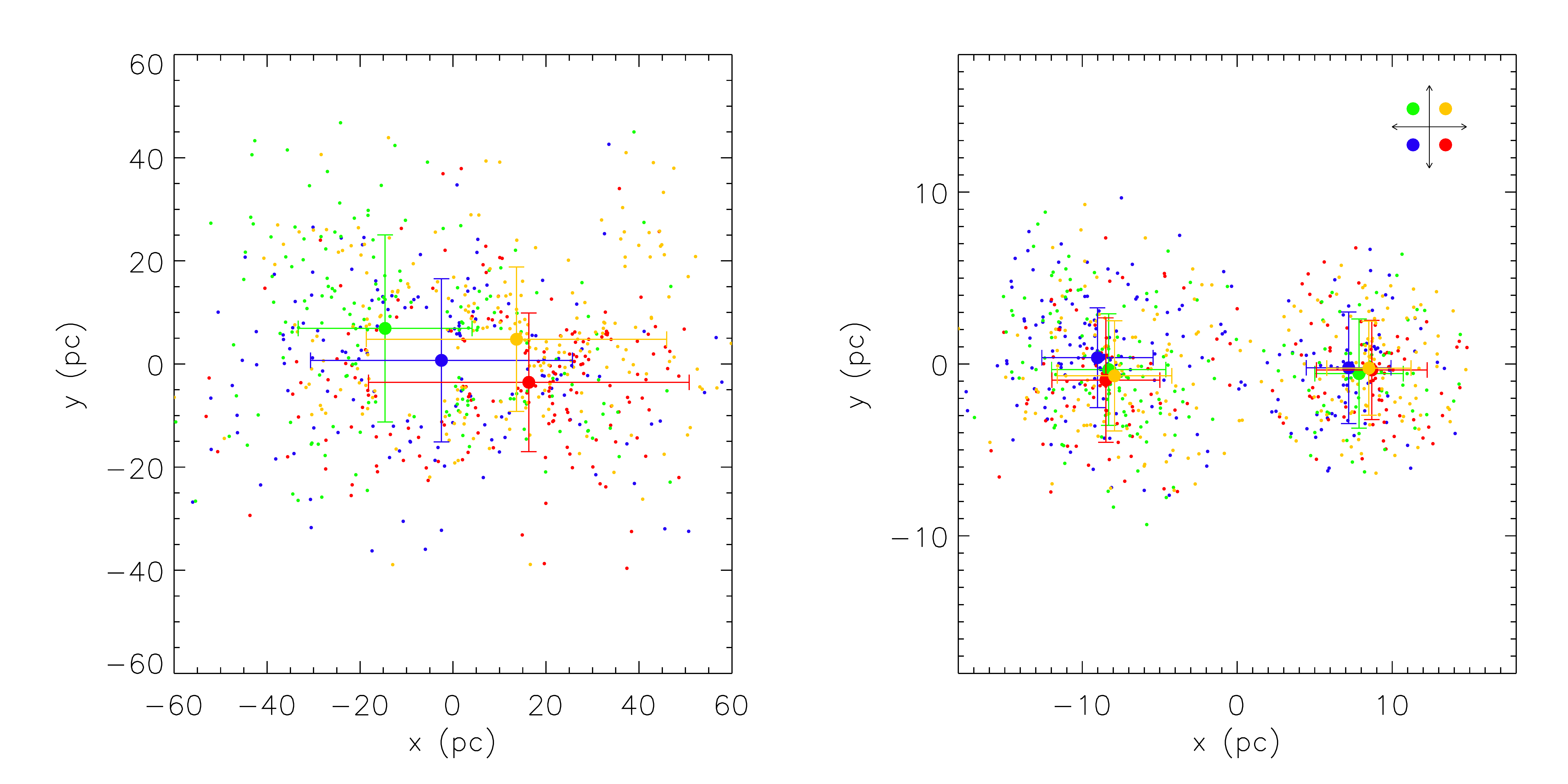}
  % \begin{minipage}[]{85mm}
   \caption{ Locations of halo (left panel) and core (right panel) components with different internal motion directions on the projected coordinate system, whose origin is located at the common center of the Double Cluster core. Colors are identical to those in figure~\ref{qmap}. The solid circles and their corresponding error bars represent the mean position and spatial dispersion of each sample, respectively. For halo and core components, the different position distribution of the four motion direction samples suggest that the tidal stripping event occurs on the outskirts of the Double Cluster, while the influence on core components is relatively small.}
%\end{minipage}
   \label{qmap_halo}
\end{figure*}

\begin{figure*}
   \centering
   \includegraphics[angle=0,scale=0.35]{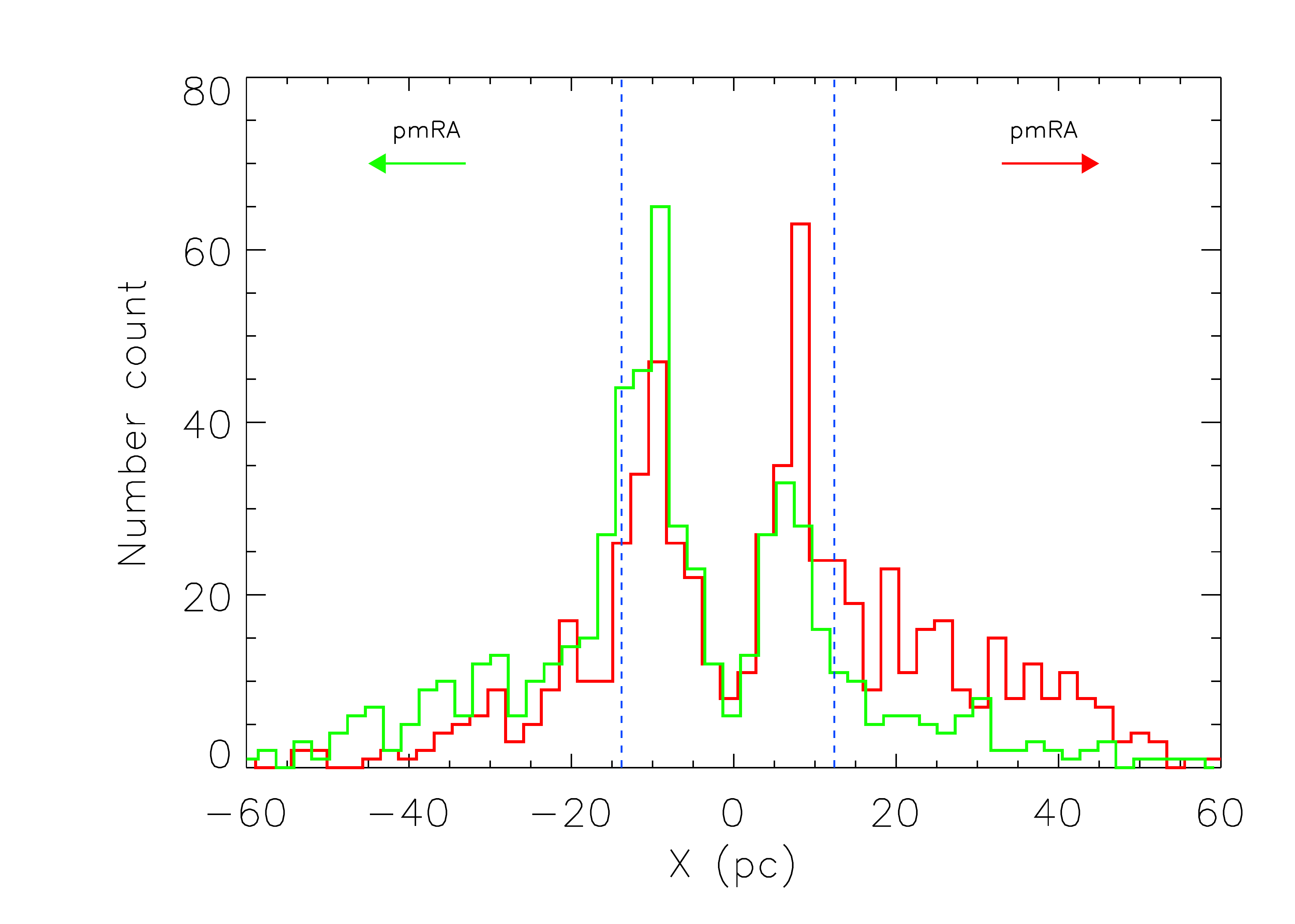}
  % \begin{minipage}[]{85mm}
   \caption{ Histograms of the different velocity groups in X coordinates. The red histograms represent stars with RA proper motion along the X direction and the green histograms represent stars with RA proper motion opposite to the X coordinates. The region between two blue dash lines mark the Double Cluster core region \citep{2013A&A...558A..53K}. It is note that, in order to show the histogram of stars belonging to cluster cores, halo stars were excluded whose position happen to locus on the core region in X coordinates (the region enclosed by the blue vertical lines).} 
%\end{minipage}
   \label{qmap_hist}
\end{figure*}

\subsubsection{Substructures}
\label{subs}
The most surprising discovery is the long-stretching filamentary substructures of the Double Cluster, which are shown in Figure~\ref{dbscan} (marked as S1 and S2). Although sub-structures have been observed in massive star-forming regions \citep{2003ARA&A..41...57L}, the scale size and extended length of cluster samples are much smaller than those of the Double Cluster, even for the linear chains of clusters classified in \citet{2014ApJ...787..107K}, whose scale length ranges from 10~pc to 30~pc. For the substructures S1 and S2 (see Figure~\ref{dbscan}), the angular radius between the common centre of the Double Cluster  and the centre of substructures are 5.1$^{\circ}$, and 2.9$^{\circ}$, corresponding to projected distances extending to 208~pc and 118~pc, respectively \citep[with an adopted cluster distance of 2344~pc;][]{2010ApJS..186..191C}.

Kinematic information can provide hints about the origin of the substructures. The relative tangential velocity between substructures and the cores of the Double Cluster are about 2~\kms (see below). This means even for the nearest substructure S2, it would take about 60~Myr to form this structure, if we assume that stars in S2 originate from the Double Cluster. This formation time greatly exceeds the Double Cluster age of 14~Myr \citep{2010ApJS..186..191C}. Therefore, these substructures are more likely to be primordial structures that have formed together with the Double Cluster core, instead of a tidally disrupted stream from the cluster center.

\subsection{Properties of Double Cluster Components}

\begin{table}
\caption{Gaussian fit results of parameters  distribution}.
\label{tab1}
\begin{tabular}{ccccc}
\hline
  & $\mu$ & $\sigma_{\mu}$ & $\varpi$ & $\sigma_{\varpi}$\\
  & mas $yr^{-1}$ & mas $yr^{-1}$ & mas & mas \\
\hline
$h$ Persei    & 1.27 &  0.13 & 0.39 & 0.04 \\
$\chi$ Persei & 1.25 &  0.12 & 0.39 & 0.04 \\
Halo          & 1.25 &  0.12 & 0.39 & 0.04 \\
S1            & 1.31 &  0.11 & 0.44 & 0.04 \\
S2            & 1.40 &  0.12 & 0.41 & 0.05 \\
\hline
\end{tabular}
\end{table}

To study the physical properties of three major Double Cluster components which labeled by the DBSCAN, including the cluster cores, the halo population, and substructures, we plot the total proper motion (top panel) and parallax (bottom panel) distribution of each component in Figure~\ref{parm}. Different colors represent different components: $h$ Persei in red, $\chi$ Persei in green, halo population in yellow, S1 in blue and S2 in violet. The Gaussian function is used to fit the histogram of each component. Fitting results of proper motion and parallax distribution are shown in Table~\ref{tab1}. The similar proper motion distribution and dispersion suggest the coeval feature of Double Cluster components. It is noted that, the parallax fitting results show the line of sight distance between core components and  substructures are about 291~pc and 125~pc for S1 and S2 respectively, which are comparable with the projected distances (see Section~\ref{subs}). 

Furthermore, it is important to constrain the age of stars belonging to different components. Figure~\ref{parm_hr} shows the isochrone fitting result for three major components, the same color-marks are used as in Figure~\ref{parm}. Referring to the literature estimation of age of the Double Cluster \citep{2010ApJS..186..191C}, we adopt the 14 Myr isochrone ($\log [t/{\rm yr}]=7.14$) with solar abundance from Padova stellar evolution models to perform the visual fit.

To transform the observational magnitude $G$mag and colors $G_{BP}-G_{RP}$ of each star to the absolute magnitude M$_{G}$ and intrinsic colors $(G_{BP}-G_{RP})_{0}$, distance modulus and reddening need to be determined. For each component, the \gaia{} DR2 provide reliable parallax measurement, the parallax of stars are used to derive their individual distance modulus. The only free parameter to adjust the isochrone fitting is the reddening $E(B-V)$. The extinction $A_G$ and reddening $E(BP-RP)$ are calculated as $A_G$=$R_G \times E(B-V)$ and $E(BP-RP)=(R_{BP}-R_{RP}) \times E(B-V)$, where $R_G=2.740$, $R_{BP}=3.374$ and $R_{RP}=2.035$, respectively \citep{2018MNRAS.479L.102C}. In our work, the best fitting isochrone derive the average $E(B-V)=0.65$ for three major components, which is slightly higher than the results of  \citep{2010ApJS..186..191C,2002ApJ...576..880S}. However, considering the dispersion of reddening for different spectral type, our result are still within a reasonable range.  

\subsection{Internal kinematic distribution}

Most of the stars in our member sample have a brightness in the range of $G = 13-17$~mag, and none of the members has radial velocity measurement in \gaia{} DR2. In this work, we therefore only use  proper motion parameters to perform the kinematic analysis.

In our sample of Double Cluster members, the measurement uncertainty of proper motion is small, with mean uncertainties $0.08 \pm 0.03$~\masyr{} in pmRA and $0.09 \pm 0.03$~\masyr{} in pmDE. Assuming all member stars have nearly the same distance, i.e., $d\approx$2.3 kpc \citep{2010ApJS..186..191C}, the tangential velocities of our sample stars range from  $-10.7$~\kms{} to $-5.2$~\kms{} in RA and $-15.2$~\kms{} to $-9.8$~\kms{} in DEC, and the mean values are $-7.7 \pm 1.2$~\kms{} in RA and $-12.2 \pm 1.2$~\kms{} in DEC, respectively.

To study the stellar relative motions in the tangential direction, especially for the internal motions within the core, halo and substructure populations,  we assume the bulk motions of core members as the zero point and  estimate the average tangential velocity of core members first. In our sample, stars within radius of 0.24$^{\circ}$ to the $h$ Persei center and radius of 0.175$^{\circ}$ to the $\chi$ Persei center \citep{2013A&A...558A..53K} are defined as core members. The average tangential velocity of core members is $-7.5 \pm 1.1$~\kms{} in RA and $-12.3 \pm 1.1$~\kms{} in DEC, nearly equal to the mean tangential velocity of all members. We then subtract the average tangential velocity from the all member stars. The residual velocity values can be consider as internal motions in the core region.

To distinguish the relative tangential velocity of different populations, we study the relative motions of member stars. For simplicity, we categorize cluster members into four quadrant samples, according to their proper motion directions. We plot a map with the projected coordinate system to show the spatial distribution of member stars using different colors, which represent their relative velocity directions belonging to different quadrants. To make the populations more distinguishable, the most crowded core regions were excluded in Figure~\ref{qmap}.

In Figure~\ref{qmap}, for stars in S1 and S2 substructure areas, we select stars within the same quadrant sample to calculate the average tangential velocities as 2.2$\pm$0.6~\kms{} and 1.9$\pm$0.7~\kms{} for S1 and S2, respectively. Here, the arrows indicate the average velocity directions. Similarly, the averaged velocity direction (angles in degrees, measured counterclockwise from the positive X coordinate) of the S1 and S2 subsamples are 148$^\circ$ $\pm$ 25$^\circ$ and 227$^\circ$ $\pm$ 25$^\circ$, respectively. 

In contrast to the bulk movements of the substructures, the tangential velocity distribution of the halo members and core members is more complex. For different velocity samples, we calculate their mean spatial position and dispersion. Figure~\ref{qmap_halo} shows the projection map of core and halo members. The colors represent different velocity samples as shown in Figure~\ref{qmap}, and the solid circles and the error bars represent the mean position and the position dispersion of stars in each sample.  The spatial distribution of halo stars in the different samples is shown in the left panel of Figure~\ref{qmap_halo}. The offset of the mean positions for each sample shows that stars with velocity in the first and fourth quadrants tend to be located in the positive region. In contrast, stars with velocities in the second and third quadrants are relatively located to the negative region. It seems that, exhibiting a bulk motion away from the cluster cores, the structure of the halo star population is experiencing certain dynamic stretching effect in the  X (RA) direction .

On the other hand, in the core region, the spatial distributions of the four velocity samples are similar and have no inclined direction, as is shown in the bottom panel of Figure~\ref{qmap_halo}. For member stars in the core region, the tangential velocity dispersion is 0.6~\kms{} for both $h$ and $\chi$ Persei. Such a small intrinsic velocity dispersion suggest the existence of dynamically compact cores in the Double Cluster.

The clusters $h$~and $\chi$~Persei have masses of roughly $5500~M_\odot$ and $4300~M_\odot$, respectively \citep{2005AJ....130..134B}, and the corresponding  tidal radii are (depending on the exact masses) roughly $23$~pc. The half-mass relaxation time of both clusters is of order 100~Myr \citep{2005AJ....130..134B}, which is substantially longer than the age of either of the star clusters. Consequently, effects of two-body relaxation and subsequent stellar ejections from the cluster cores is limited. We carry out $N$-body simulations using {\tt NBODY6++GPU} \citep{wang2015, wang2016} and find that only a hand-full of stars (primarily neutron stars that experienced a velocity kick) are ejected beyond the tidal radius within the age of the star clusters, while the effect of the Galactic external tidal field does not result escape through evaporation on this timescale. Conversely, a much larger number of stars found in the Double Star cluster halo. The different dynamical properties of core and halo population imply that the tidal stripping event only had a significant influence on the outskirts of the Double Cluster, while the impact on the core components is relatively small.

To further clarify the intrinsic velocity offset in the X coordinate, we classify halo and core members into two velocity groups: a red group with positive RA proper motions and a green group with negative RA proper motions. The projected spatial distribution of the two velocity groups are shown in Figure~\ref{qmap_hist}.  The distribution pattern in Figure~\ref{qmap_hist} also demonstrates that stars with positive RA proper motions tend to be located in the positive region in X coordinate, while stars with negative RA proper motions tend to be located in the negative region in X coordinate.

\section{Summary}

We have studied the extended spatial morphology and kinematic properties of the Double Cluster $h$ and $\chi$ Persei, using unprecedented high-precision data from the \gaia{} DR2. As suggested by \citet{2010ApJS..186..191C}, an extended halo structure has been found at an angular distance beyond one degree from the Double Cluster core. Furthermore, a long-stretching filamentary substructure is discovered for the first time, which extend to a projected distance of more than 200~pc away from the Double Cluster centre. These extended structures suggest that cluster formation mode and its history may be more complex than previously expected.

Based on our internal kinematic analysis, the small tangential velocity between substructure and cluster core implies that filamentary substructures are more likely to be primordial structures, reminiscent from the star formation event that formed the Double Cluster. In addition,  for members in the halo region, the tangential velocity distribution suggests that a tidal stripping event is occurring in the outskirts of the Double Cluster. By contrast, core components of the Double Cluster still maintain the dynamical compact properties.

Although the \gaia{} spectroscopic data is absent in our kinematic analysis, the ongoing \lamost{}-MRS survey (LAMOST Collaboration et al. in preparation) is arranged to observe a large number of member sources in the Double Cluster field within a 5~degree diameter. More than 20\,000 median-resolution stellar spectra ($R \sim 7500$, $\sigma_{RV} \sim 1-2$~\kms{}) are expected to be provided by the \lamost{}-MRS survey, while the completeness of cluster member stars ($bp$mag $\leq$ 15 mag) can reach to about 80\%. After combining with the \lamost{} data and \gaia{} data, the three-dimensional internal kinematic properties of cluster components as well as the metallicity distribution can be studied in further detail.

\begin{acknowledgements}
The authors acknowledges the National Natural Science Foundation of China (NSFC) under grants U1731129 and 11503066 (PI: Zhong), 11373054 and 11661161016 (PI: Chen), 11390373 (PI: Shao), M.B.N.K. acknowledges support from the National Natural Science Foundation of China (grant 11573004) and the Research Development Fund (grant RDF-16-01-16) of Xi'an Jiaotong-Liverpool University (XJTLU).
This project was developed in part at the 2018 \gaia{}-\lamost{} Sprint workshop, supported by the National Natural Science Foundation of China (NSFC) under grants 11333003.
This work has made use of data from the European Space Agency (ESA) mission \gaia{} (\url{https://www.cosmos.esa.int/gaia}), processed by the \gaia{} Data Processing and Analysis Consortium (DPAC,\url{https://www.cosmos.esa.int/web/gaia/dpac/consortium}). Funding for the DPAC has been provided by national institutions, in particular the institutions participating in the \gaia{} Multilateral Agreement.
\end{acknowledgements}


\begin{thebibliography}{}

\bibitem[Bailer-Jones et al.(2018)]{2018AJ....156...58B} Bailer-Jones, C.~A.~L., Rybizki, J., Fouesneau, M., Mantelet, G., \& Andrae, R.\ 2018, \aj, 156, 58 

\bibitem[Bragg \& Kenyon(2005)]{2005AJ....130..134B} Bragg, A.~E., \& Kenyon, S.~J.\ 2005, \aj, 130, 134

\bibitem[Bragg \& Kenyon(2002)]{2002AJ....124.3289B} Bragg, A.~E., \& Kenyon, S.~J.\ 2002, \aj, 124, 3289 

\bibitem[Cantat-Gaudin et al.(2018)]{2018A&A...618A..93C} Cantat-Gaudin, T., Jordi, C., Vallenari, A., et al.\ 2018, \aap, 618, A93.

\bibitem[Casagrande \& VandenBerg(2018)]{2018MNRAS.479L.102C} Casagrande, L., \& VandenBerg, D.~A.\ 2018, \mnras, 479, L102 

\bibitem[Castro-Ginard et al.(2018)]{2018A&A...618A..59C} Castro-Ginard, A., Jordi, C., Luri, X., et al.\ 2018, \aap, 618, A59.

\bibitem[Currie et al.(2007)]{2007ApJ...659..599C} Currie, T., Balog, Z., Kenyon, S.~J., et al.\ 2007, \apj, 659, 599

\bibitem[Currie et al.(2010)]{2010ApJS..186..191C} Currie, T., Hernandez, J., Irwin, J., et al.\ 2010, \apjs, 186, 191

\bibitem[Dias et al.(2002)]{2002A&A...389..871D} Dias, W.~S., Alessi, B.~S., Moitinho, A., \& L{\'e}pine, J.~R.~D.\ 2002, \aap, 389, 871

\bibitem[Evans et al.(2009)]{2009ApJS..181..321E} Evans, N.~J., II, Dunham, M.~M., J{\o}rgensen, J.~K., et al.\ 2009, \apjs, 181, 321

\bibitem[Faesi et al.(2016)]{2016ApJ...821..125F} Faesi, C.~M., Lada, C.~J., \& Forbrich, J.\ 2016, \apj, 821, 125

\bibitem[Gaia Collaboration et al.(2016)]{2016A&A...595A...1G} Gaia Collaboration, Prusti, T., de Bruijne, J.~H.~J., et al.\ 2016, \aap, 595, A1

\bibitem[Gaia Collaboration et al.(2018)]{2018A&A...616A...1G} Gaia Collaboration, Brown, A.~G.~A., Vallenari, A., et al.\ 2018, \aap, 616, A1.

\bibitem[Garmany \& Stencel(1992)]{1992A&AS...94..211G} Garmany, C.~D., \& Stencel, R.~E.\ 1992, \aaps, 94, 211

\bibitem[Heyer \& Dame (2015)]{2015ARA&A..53..583H} Heyer, M., \& Dame, T.~M.\ 2015, \araa, 53, 583

\bibitem[Kharchenko et al.(2005)]{2005A&A...438.1163K} Kharchenko, N.~V., Piskunov, A.~E., R{\"o}ser, S., Schilbach, E., \& Scholz, R.-D.\ 2005, \aap, 438, 1163

\bibitem[Kharchenko et al.(2013)]{2013A&A...558A..53K} Kharchenko, N.~V., Piskunov, A.~E., Schilbach, E., R{\"o}ser, S., \& Scholz, R.-D.\ 2013, \aap, 558, A53


\bibitem[Kuhn et al.(2014)]{2014ApJ...787..107K} Kuhn, M.~A., Feigelson, E.~D., Getman, K.~V., et al.\ 2014, \apj, 787, 107

\bibitem[Lada et al.(1993)]{1993prpl.conf..245L} Lada, E.~A., Strom, K.~M., \& Myers, P.~C.\ 1993, Protostars and Planets III, 245

\bibitem[Lada \& Lada(2003)]{2003ARA&A..41...57L} Lada, C.~J., \& Lada, E.~A.\ 2003, \araa, 41, 57

\bibitem[Luo et al.(2015)]{2015RAA....15.1095L} Luo, A.-L., Zhao, Y.-H., Zhao, G., et al.\ 2015, Research in Astronomy and Astrophysics, 15, 1095

\bibitem[Marsh Boyer et al.(2012)]{2012AJ....144..158M} Marsh Boyer, A.~N., McSwain, M.~V., Aragona, C., \& Ou-Yang, B.\ 2012, \aj, 144, 158

\bibitem[Oosterhoff(1937)]{1937AnLei..17A...1O} Oosterhoff, P.~T.\ 1937, Annalen van de Sterrewacht te Leiden, 17, A1

\bibitem[Portegies Zwart et al.(2010)]{2010ARA&A..48..431P} Portegies Zwart, S.~F., McMillan, S.~L.~W., \& Gieles, M.\ 2010, \araa, 48, 431

\bibitem[Priyatikanto et al.(2016)]{2016MNRAS.457.1339P} Priyatikanto, R., Kouwenhoven, M.~B.~N., Arifyanto, M.~I., Wulandari, H.~R.~T., \& Siregar, S.\ 2016, \mnras, 457, 1339

\bibitem[Rastorguev et al.(1999)]{1999AstL...25..595R} Rastorguev, A.~S., Glushkova, E.~V., Dambis, A.~K., \& Zabolotskikh, M.~V.\ 1999, Astronomy Letters, 25, 595

\bibitem[Slesnick et al.(2002)]{2002ApJ...576..880S} Slesnick, C.~L., Hillenbrand, L.~A., \& Massey, P.\ 2002, \apj, 576, 880

\bibitem[Uribe et al.(2002)]{2002PASP..114..233U} Uribe, A., Garc{\'{\i}}a-Varela, J.-A., Sabogal-Mart{\'{\i}}nez, B.-E., Higuera G., M.~A., \& Brieva, E.\ 2002, \pasp, 114, 233

\bibitem[Xiang et al.(2015)]{2015MNRAS.448..822X} Xiang, M.~S., Liu, X.~W., Yuan, H.~B., et al.\ 2015, \mnras, 448, 822

\bibitem[Wang et al.(2015)]{wang2015} Wang, L., Spurzem, R., Aarseth, S., et al.\ 2015, \mnras, 450, 4070 

\bibitem[Wang et al.(2016)]{wang2016} Wang, L., Spurzem, R., Aarseth, S., et al.\ 2016, \mnras, 458, 1450 

\bibitem[Zhao et al.(2012)]{2012RAA....12..723Z} Zhao, G., Zhao, Y.-H., Chu, Y.-Q., Jing, Y.-P., \& Deng, L.-C.\ 2012, Research in Astronomy and Astrophysics, 12, 723

\end{thebibliography}
\end{document}